# Unveiling High-Pressure Investigation of BeX (X = S, Se, and Te): A DFT- Base Exploration of Phonon Spectra, Molecular Dynamics, Optical Responses, and Thermodynamic stability for Advance Optoelectronic Applications

Muhammad Shahzad [a], Sikander Azam[b], Syed Awais Ahmad,[c] Ming Li [a] [*]

[a] *Solar Energy Research Institute Yunnan Normal University Kunming Yunnan 650500 P.R. China*

[b] *Faculty of Engineering and Applied Sciences, Riphah International University Islamabad, 44000, Pakistan.*

[c] *Yunnan Key Laboratory of Opto-Electronic Information Technology, School of Physics and Electronics Information, Yunnan Normal University, Kunming Yunnan 650500, P.R.China*

**Corresponding author:** (Ming Li) lmllldy@126.com

## Abstract

In this study, we focus on the structure, electronic, optical and thermodynamic properties of BeX (X = S, Se, Te) under hydrostatic pressure changes from 0 to 10 GPa. The computations were made through the Generalized Gradient approximation (GGA) and Perdew-Burke-Ernzerhof functional (PBE) utilizing the CASTEP code. It was demonstrated through phonon dispersion studies that the three compounds maintain their dynamic stability at all applied pressures because imaginary frequencies were absent everywhere in the Brillouin zone. Our study revealed that pressure puts stress on all the materials studied and BeS still maintains the prevalent electronic bandgap. Different optical properties such as dielectric functions, absorption spectra, reflectivity and energy loss, were studied in detail for photon energies less than 30 eV. Analysis of optical absorption spectra indicates significant optical activity with maximum photon absorption occurring in UV region. Furthermore, thermodynamic properties like Debye temperature, heat capacity and entropy were studied. When the pressure goes up, atoms move less and therefore heat capacity decreases. When there is constant pressure, the slope of the Gibbs free energy curve tilts slightly greater which reveals a steady variation of entropy with temperature. The findings confirm that BeX (X = S, Se, Te) has enhance thermodynamic properties and Suggest promising applications in optoelectronics, thermoelectric and thermal barriers, especially in pressure dependent optoelectronic devices.

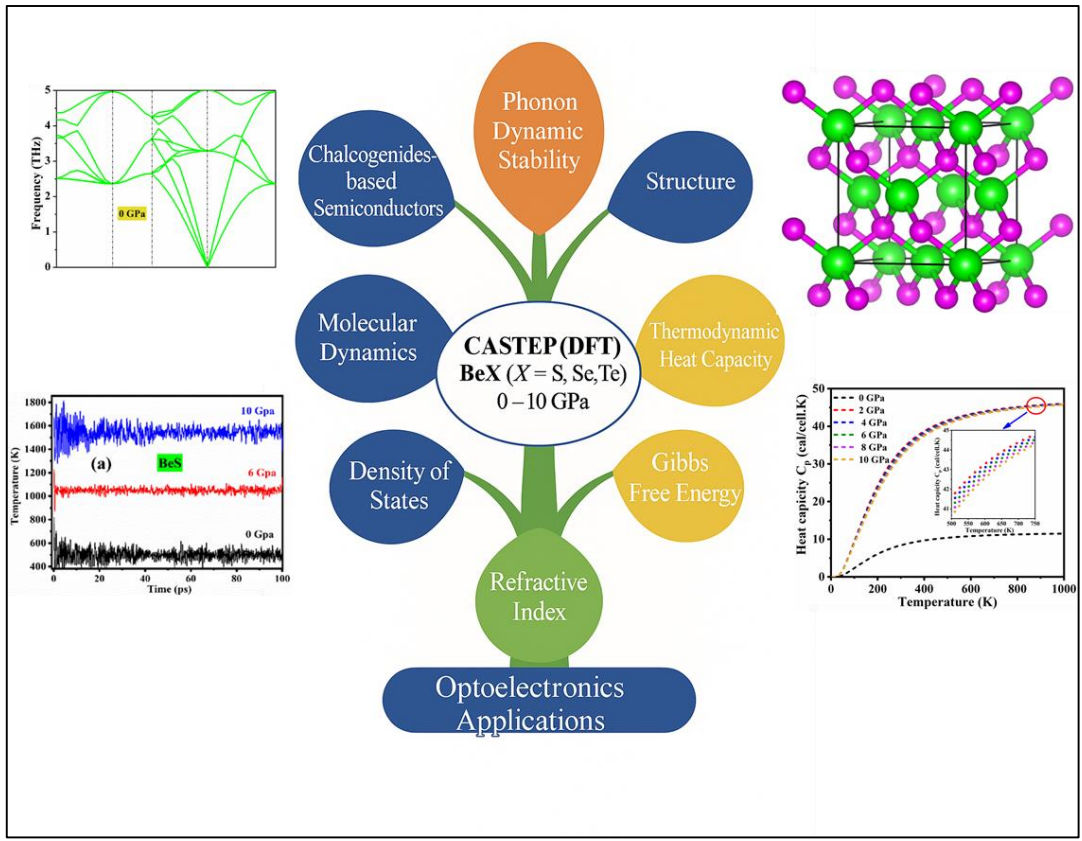

**Graphical abstract**

## 1. Introduction

The global push for environmentally friendly sustainable energy systems increased research into future-oriented technology materials. The research community strongly focuses on solar energy because it represents the most sustainable renewable power source requiring efficient capture and conversion methods. The research community has recognized chalcogenides-based semiconductors as a leading class of materials due to their promising properties for photovoltaic technologies, optoelectronic devices, energy storage systems, and catalytic applications [1, 2].

Beryllium-based semiconductors are a promising but relatively underexplored category of materials in electronic and optoelectronic devices engineering. These materials, such as beryllium Sulfide (BeS), beryllium selenide (BeSe) and beryllium telluride (BeTe) belonging to II-IV group of elements are known for their unique electronic properties when subjected to pressure, including high carrier mobility, wide bandgap, and excellent thermal conductivity. They have garnered increasing interest for use in high-frequency, high-power, and radiation-hardened devices due to their potential advantages over traditional semiconductors like silicon and gallium arsenide. However, challenges related to material synthesis, dynamic stability, and integration into existing manufacturing processes pose significant obstacles to their widespread use [3, 4]. In the stable phase, these materials exhibit indirect bandgap. Notably, these compounds exhibit a higher amount of covalent bonding as compared to group II-VI materials, with bond strengths of BeS, BeSe and BeTe being similar to those seen in GaN [5].

Research interest in Beryllium chalcogenides BeS, BeSe and BeTe continues to grow because of their appealing properties which is suitable for microelectronic, optoelectronic and spintronic device applications. These materials display advantageous electronic properties but exist as nonmagnetic substances because researchers have not extensively explored magnetic integration [6]. The beryllium chalcogenides are highly toxic [7] which has restricted the possible experiments work. For this reason, there are only a few studies that experimentally test these ideas in published literature[8, 9] . Some articles have been found on ternary compounds such as $BeS_xSe_{1-x}$ $BeS_xTe_{1-x}$, and $BeSe_xTe_{1-x}$ ($0 \leq x \leq 1$). Many studies have focused on the structure, electronics, elasticity, and optical features of Be-chalcogenides. Band gap engineering for $BeS_xSe_{1-x}$, $BeS_xTe_{1-x}$, and $BeSe_xTe_{1-x}$ (x = 0, 0.5, 1) was done with density functional theory (DFT), however, the conclusion about whether the gap is direct or indirect was inconclusive, which left some questions about their usefulness in optoelectronic applications. Besides, studies that use the virtual crystal approximation and the tight-binding model found out what happens to the elastic properties and optical features when pressure is applied to various combinations of these ternary substances [10]. Numerous theoretical studies have been carried out to explore the magnetic, electronic, and structural properties of BeX (X = S, Se, Te) compounds doped with various transition metals. These investigations often report the emergence of half-metallic ferromagnetic (HMF) behavior in transition metal-doped BeX systems. In recent years, first-principles calculations involving dopants such as V, Co, N, Cr, and Mn have been presented, highlighting their impact on the physical characteristics of BeX materials [11, 12]. Among them, BeS stands out for its exceptional mechanical stability and rigidity, attributed to its high bulk modulus and wide band gap, whereas BeTe exhibits semiconducting behavior with a comparatively narrower energy gap. The energy band gap, among other factors, plays a crucial role in determining the overall physical properties of semiconductors [13]. By employing different pressure, it is possible to tailor these properties such as thermal and dynamic stability or reducing the band gap to meet specific application requirements.

In this work we highlight that beryllium chalcogenides are wide-bandgap II-VI semiconductors with distinct structural and electronic properties that distinguish them from other well-studied chalcogenides. Despite their potential for optoelectronic applications (such as in high-energy devices, photo-detectors, and transparent electronics), systematic studies of BeS, BeSe, and BeTe remain very limited in the literature. In particular, there is a lack of detailed investigations into their thermodynamic stability, electronic density of states, and pressure-dependent behavior, which represent important blank spots in the current research landscape [14].

The innovative aspect of our work lies in providing a comprehensive theoretical study of these materials using ab initio molecular dynamics simulations. By analyzing their structural, energetic, and electronic responses under varying pressure conditions, we contribute new insights that can guide future experimental and applied research on these compounds.

## 2. Computational details

The cubic F-43m BeX (where X = S, Se, Te) compounds were thoroughly investigated using the Cambridge Serial Total Energy Package (CASTEP) code [15, 16], a highly sophisticated tool for performing first-principles calculations based on Density Functional Theory (DFT). This software applies the principles of quantum mechanics to predict material properties with high accuracy. The Perdew-Burke-Ernzerhof (PBE) exchange-correlation functional [17, 18], within the framework of the Generalized Gradient Approximation (GGA), was employed to handle the exchange and correlation effects between electrons. The principal quantum numbers For BeS, the 2s orbital of beryllium (Be) and the 2p orbital of sulfur (S) are the main contributors to the electronic structure, which are reflected in the DOS diagram as 2s (Be) and 2p (S), respectively. In BeSe, the principal quantum numbers for beryllium (Be) remain the same, with the 2s orbital labeled as 2s (Be). However, selenium (Se) contributes through its 3p orbitals, which are represented as 3p (Se) in the DOS diagram. For BeTe, the 2s orbital of beryllium (Be) again plays a key role, labeled as 2s (Be), while tellurium (Te) contributes through its 4p orbitals, labeled as 4p (Te) . These labels in the DOS diagrams provide a clear indication of the specific orbitals contributing to the electronic structure in each material. This approach allows for the efficient and precise treatment of the electronic structure of the material under study. In terms of computational techniques, the plane-wave pseudo potential method was utilized, which is a standard approach in DFT calculations. This technique approximates the interactions between electrons and atomic nuclei, replacing the complex interactions with simpler, effective potentials that are easier to compute. In particular, the ultrasoft pseudo potential method was used to model the interaction between the valence electrons and atomic cores more efficiently, thus enabling a more accurate description of the electronic structure. For the pressure-dependent analysis, the calculations were performed at an initial pressure of 0 GPa, representing the material in its equilibrium state. Subsequently, the pressure was gradually increased to 10 GPa to study the material's behavior under compression. To reach the lowest possible energy configuration under external pressure, both the atomic positions and lattice vectors were optimized. This ensures that the system achieves its minimum-energy state (equilibrium), which is crucial for understanding material properties under different conditions. The TPDS (Temperature Pressure and Stress Dependent Structure) reduction technique was applied to relax the structure. This technique is a well-established method for relaxing atomic configurations by considering the effects of both pressure and temperature, which is important for obtaining a realistic structural model. The calculations converged with specific tolerance values: 0.05 GPa for pressure, $1 \times 10^{-6}$ eV/atom for energy, and 0.04 eV/Å for force, ensuring high precision in the simulation results. A cutoff energy of 500 eV was selected for the plane-wave basis set, which ensures that the expansion of the wave-functions in terms of plane waves is sufficiently accurate while maintaining computational efficiency. Additionally, a k-point grid of 11 × 11 × 11 inside the unit cell was used to sample the Brillion zone, ensuring accurate integration of the electronic structure over the entire momentum space. To investigate the dynamic stability and thermodynamic properties of the BeX (X = S, Se, Te) compounds under varying pressures, phonon calculations were performed. Phonon calculations provide valuable information about the material's vibrational modes and help assess its stability by evaluating the frequency of these vibrations, which are crucial for understanding the material's response to thermal and mechanical stresses.

Overall, these computational techniques, implemented in CASTEP, provide a comprehensive theoretical framework for studying the structural, electronic, optical and thermodynamic properties of the BeX (X = S, Se, Te) compounds under the effect of different pressure shedding light on their behavior under different environmental conditions.

## 3. Results and discussions
### 3.1. Structural properties

Depending on the halide (S, Se and Te), the BeX (X = S, Se and Te) compounds have a crystalline structure that cubic in type. The BeS is commonly forms in a zinc blende F-43m or wurtzite structure and same for other structure in the group which is based on appropriate replacements. Figure 1 shows how $Be^{2+}$ atoms in the unit cell are connected to four $S^{-2}$ atoms on all four corners of a tetrahedron. Usually, a unit cell groups have 5 atoms and 1 Be surrounded by 4 S, Se and Te atoms in a tetrahedral structure, but this arrangement and symmetry vary when the S, Se and Te atom type is different. There bond length throughout the crystal lattice, when the unit cell is put under pressure, its overall volume decreases and it is true for many covalent or partially ionic compounds. The study aims to examine the effects of pressure on the structure of BeX (X = S, Se and Te) compounds in a systematic manner. In figure 1, the crystal structure of BeX (X = S, Se and Te) is shown and the Be with S, Se and Te atoms are arranged following the rules for the space group. Depending on the setup, Be could be located in either the 2b or 1a position and S, Se and Te atoms could go to 2c or 4f. The calculated characteristics of the structure while under pressure are shown in Table 1.

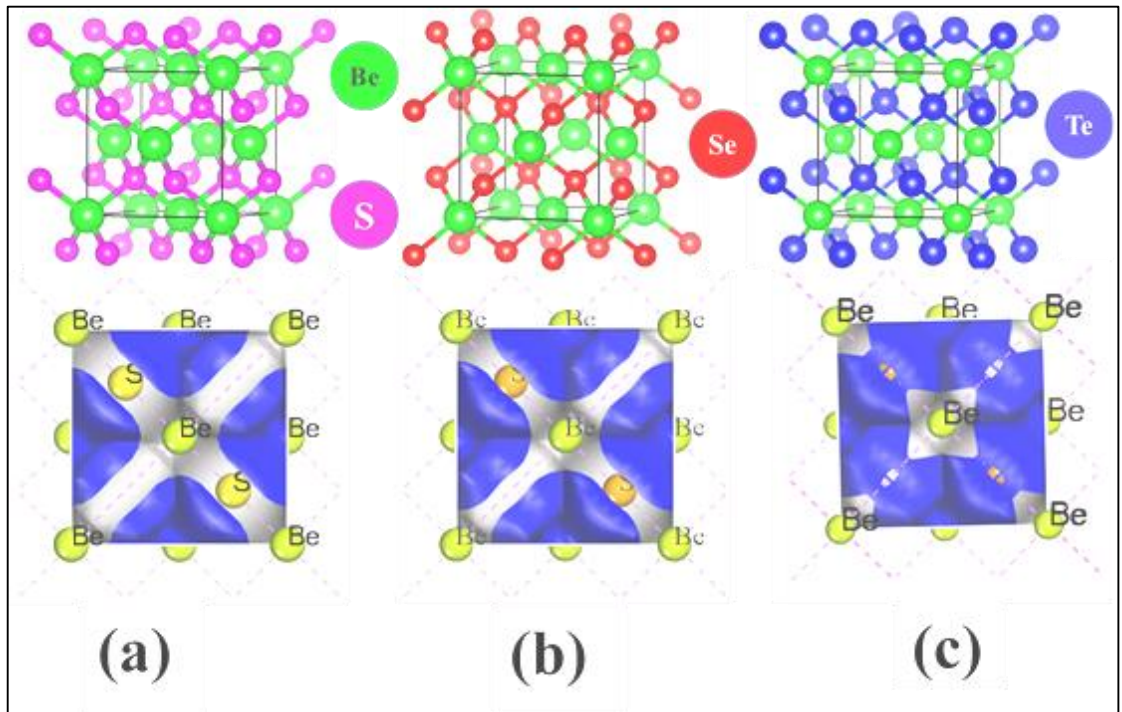

**Fig.1**. Crystalline structure and electron density distribution of (a) BeS (b) BeSe and (c) BeTe

There is a good correspondence between these theoretical data and the experiments, meaning that the phase has not changed much during the surveyed pressure range. From this, we can see that crystal symmetry and structure are not significantly affected by high or moderate pressure within BeX (X = S, Se and Te) compounds. The graphs in Figure 2(a-b) illustrate the pressure-dependent behavior of lattice constants and BeX (X = S, Se and Te) bond lengths. Both parameters show a negative correlation with increasing pressure, indicating that the structure becomes more compact as pressure rises from 0 to 10 GPa. This compressive behavior may significantly influence the electronic, optical thermodynamic properties of BeX (X = S, Se and Te) materials, highlighting their potential for pressure-tuned applications.

The electrons density in figure 1 shows how electrons are distributed in BeX (X = S, Se and Te) around the atoms and within the bond of crystal structure. High electron density around S, Se and Te atoms indicating they attract more electrons with higher electronegativity. The low density shows the atom lose the electrons and form cations while moderate density indicating partial covelncy and electron sharing to ionic transfer.

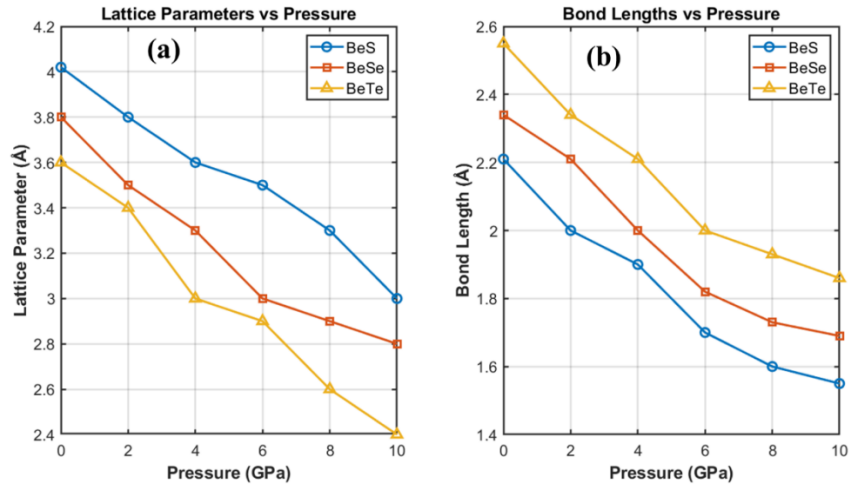

**Fig.2.** The estimated (a) lattice parameters and (b) bond lengths of cubic BeX (X = S, Se and Te) under the applied external pressure varies 0-10 GPa.

### 3.1.1. Dynamic stability

To understand the vibrations and thermodynamic behavior of the compounds, their phonon dispersion relations were examined. Phonon dispersion reveals how vibrations in an atom influence key aspects of the material such as heat transfer, the total heat, entropy and the stability of its structure [19, 20].

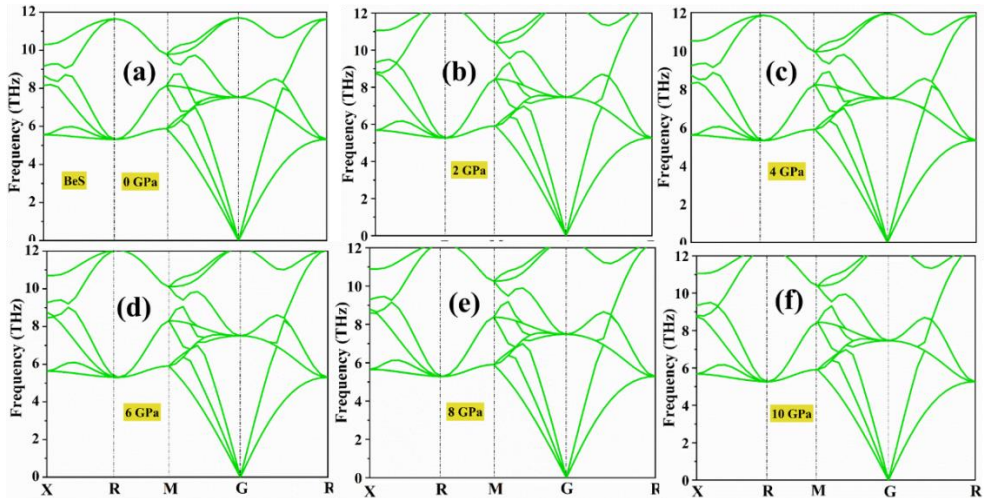

**Fig.3.** phonon curve of BeS under varying pressure

We performed phonon calculations using CASTEP [14] which is a part of the Materials Studio software. It depends on density functional perturbation theory (DFPT) which effectively explains the reaction of atoms to minor changes in positions. A map of the phonon band structure was made by evaluating the dynamical matrix at important points in the Brillion zone. A phonon dispersion curve free from imaginary frequencies (no negative modes) serves as a strong indicator of dynamical stability of the crystal structure. Conversely, the presence of negative frequencies may suggest structural instabilities or possible phase transitions under certain conditions, such as temperature or pressure.

This methodology enables us to comprehensively assess the structural integrity and lattice dynamics of the compounds, which are critical for determining their suitability for applications in thermal management, optoelectronics, or high-pressure environments.

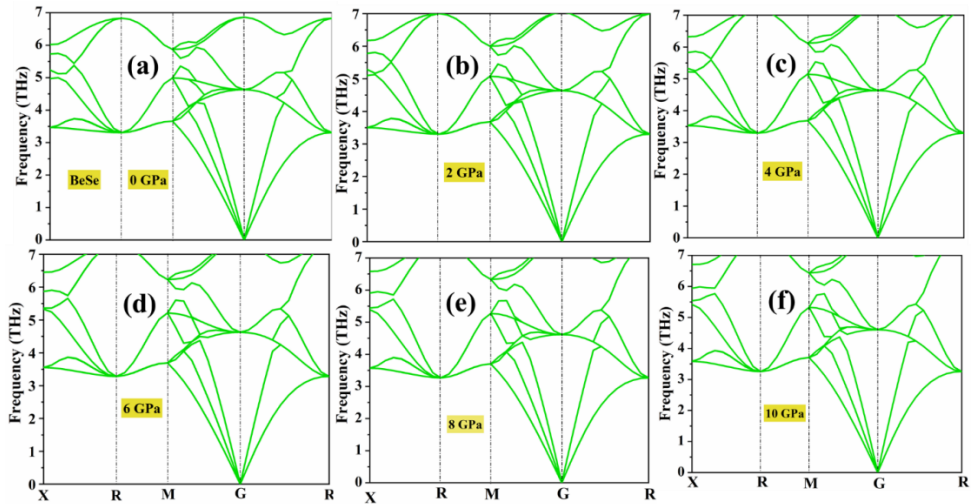

**Fig.4.** phonon curve of BeSe under varying pressure

Here, we used detailed phonon dispersion tools to find out the dynamically stability of BeX (X = S, Se and Te) under the applied external pressure. Phonons are related to the movements of atoms in a solid material. If only real and positive phonons are present in the Brillouin zone, this means that the material can resist small variations and remain stable in its structure [21]. In some cases, when there are imaginary phonons on the dispersion graphs (plotted as negatives), it indicates that the lattice may become unstable and prompt changes in material structure, phase transitions or soft-mode effects when adjusting conditions such as temperature or pressure. The DFPT calculating phonon dispersion relations allows us to learn more about the impact of atomic interactions on the properties of the material. This is how we find out about low-energy processes which directly impact thermal conductivity, high-temperature behavior and functioning of a device.

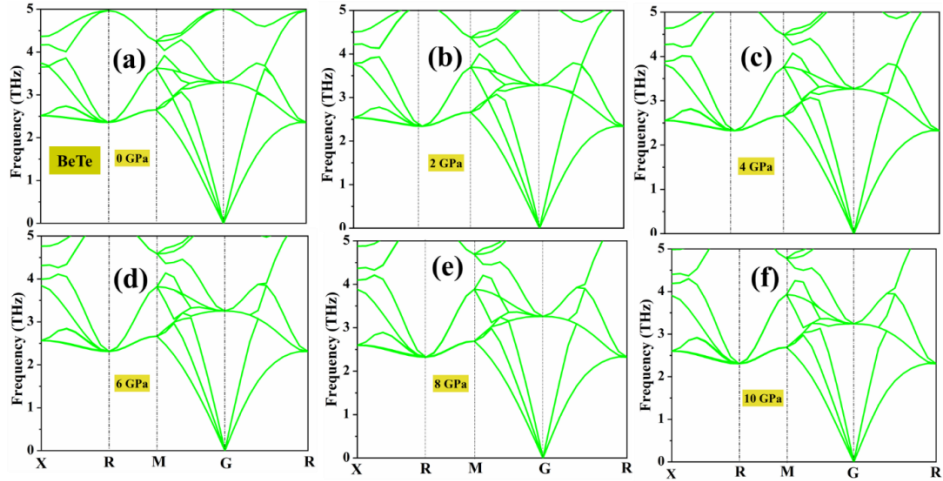

**Fig.5**. phonon dispersion curve of BeTe under varying pressure.

Figure 3, 4 and 5 shows the calculated dispersion curves for BeX (X = S, Se and Te) under pressures from 0 to 10 GPa. The examination of the phonon spectra reveals only real and positive frequencies in all the studied pressures, indicating that the materials continue to be dynamically stable. The fact that imaginary modes do not appear in the computations shows there is no risk of instability or transition. The Phonons spectra in BeX (X = S, Se and Te) compounds have three acoustic branches which result from longitudinal movements initiated by vibrations of the structure's longest wavelengths. Phonons with a zero wave-vector are found near the G center of the Brillouin zone and these are called acoustic branches. When the wave-vector rises, these modes move smoothly from one mode to the next, reflecting stable vibrations of the crystal system. The acoustic modes lead to a group of optical phonon branches appearing because the BeX (X = S, Se and Te) atoms move against each other, because of mass difference and bond characteristic. In figure 3 BeS shows higher optical phonon frequency as compared to BeSe and BeTe as shown in figure 4 and 5. These results are consistent with what is predicted by mass-dependent phonon softening, with heavier atoms vibrating less quickly. This behavior aligns with the expected mass-dependent phonon softening, where heavier atoms vibrate more slowly. The phonon gap between acoustic and optical branches may also shrink with increasing atomic mass.

### 3.1.2. Molecular dynamic simulation (MD)

The MD simulations, enables how atoms and molecules in a system move with time. By modeling tiny interactions, MD helps to understand the changes in materials influenced by strain, temperature and pressure. MD simulation is based on using Newton's second law of motion (F =

ma) for every single atom in a system. By adding up the nearby atom forces according to a potential (e.g. Lennard-Jones, EAM or Tersoff) or force field from that we get the total force acting on every atom. The forces determined, the simulation changes the locations and speeds of atoms over very small intervals of time (typically picoseconds) using different integration methods [22]. By using simulations, we can examine how changes happening at an atomic or molecular scale guide the actions of a material system. MD helps researchers explore how structural, mechanical and dynamic properties are affected by changes in atoms, defects, pressures or temperature by running simulations [23].

The process of MD simulations are applied to BeX (X = S, Se, Te) compounds to monitor their energy and temperature changes as time passes. This is investigating by following and mapping the changes in potential energy (P.E.), kinetic energy (K.E.), total energy (T.E.) and temperature with changes in simulation time. The figure. 6 shows how P.E, K.E, T.E and temperature shift with time for BeX (X = S, Se, Te) compounds at different hydrostatic pressures. During the simulation, the total energy does not change much which demonstrates that the system has found a dynamic balance. Note that P.E and K.E shows inverse trends, so a rise in one normally means a fall in the other. This relationship points out that atomic motion and interatomic forces are always competing with each other. The system becomes more stable when the movement in atoms are increases and as result interactions between the atoms are made more attractive. The fact that the atoms shows the same motion even under thermal agitation proves that they are still attached by strong interatomic links. When external pressure are applied to overall BeX (X = S, Se, Te) compounds both K.E and P.E go up in total. Higher kinetic energy means the atoms in the substance become more active and hotter and higher potential energy shows that the atoms are packed closely, strengthening both their repulsion and attraction. As a result, BeX (X = S, Se, Te) materials grow denser and more compact at high pressure which matches their important properties as high-energy, bonded compounds. The related patterns in the plot as shown in figure 6 suggest that BeS, BeSe and BeTe bonded similarly and handling pressure at the same way. Although atomic weights and bond strength cause minor differences in the results.

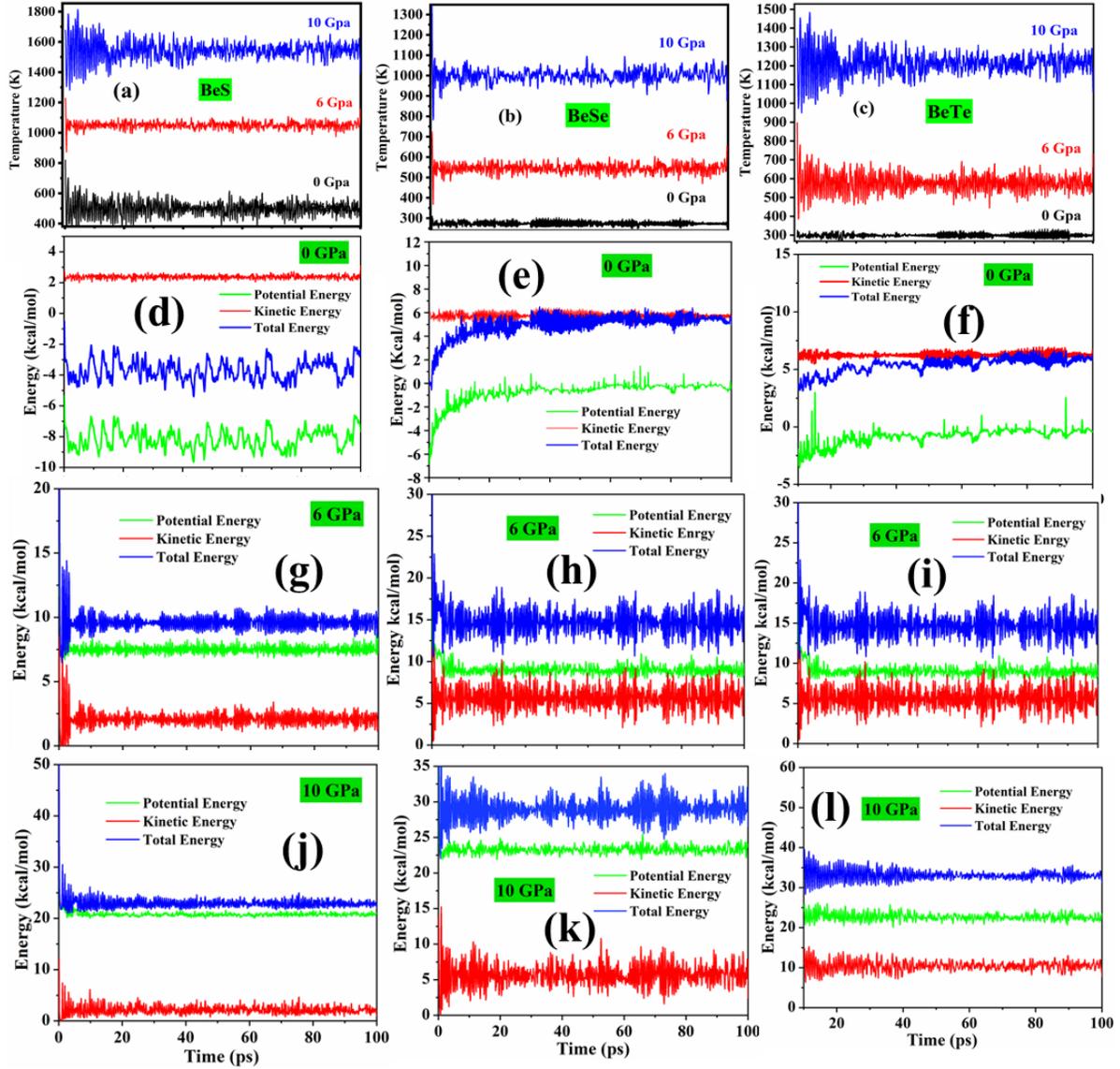

**Fig.6.** Temperature energy and evolution over simulation time steps for (a-j) BeS (b-k) BeSe (c-l) BeTe at varying pressure levels.

The changes in K.E in BeX (X = S, Se, Te) compounds during a picosecond is reflected by subtle fluctuations in temperature during the same period. Since the temperature-time graph has minor fluctuations, it doesn't seem that the system is facing any sudden changes in energy or unstable structure. They say that an energetic balance, called dynamic equilibrium which keeps the BeX (X = S, Se, Te) crystal lattice strong even during the usual atomic vibrations triggered by heat. In conclusion the results shows that BeS, BeSe and BeTe demonstrate stability as the conditions are changed. The crystal structures resist deformation, a material's ability to distribute heat without undergoing significant structural changes which confirms its mechanical strength.

This is particularly important for BeX (X = S, Se, Te) compounds, as their wide band gaps and strong ionic-covalent bonding provide high resistance to both temperature and pressure. The temperature peak values for BeS, BeSe and BeTe are found near 1700, 1100 and 1400K after about 5 ps respectively. A rapid increase in temperature means that the system is taking in heat during those areas of the trace. Applying high pressure together assistances increase the internal energy of atoms and pushed closer together, which raising both the systems kinetic and potential energy components. It confirms that when the temperature changes, the material's energy profile also varies in the same way. Because temperature and internal energy move in unison across BeX (X = S, Se, Te) compounds, they remain predictable in extreme environments. Its ability to survive in both high-pressure and high-temperature conditions depends on the strong ionic-covalent bonds and consistent structure of their crystals. The thermodynamic parameters at constant pressure are shown in figure 7.

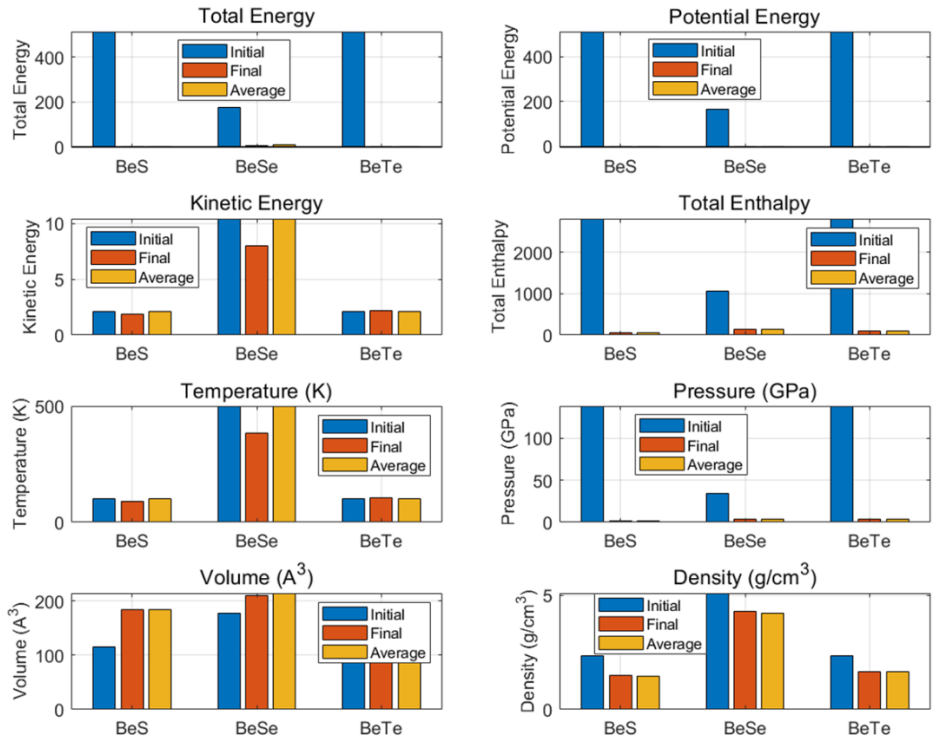

**Fig.7.** Dynamic parameters comparison of BeX (X = S, Se, Te) at constant pressure.

## 3.2. Electronic properties
### 3.2.1. Electronic band structure

The electronic band structure describes energy levels of electrons within a solid. To learn how materials conduct electricity, respond to light and behave in electronic purposes is highly important. A crystal lattice is made by atoms grouping together in a regular pattern. If such conditions exist, electrons inside isolated atoms have their energy levels combine and expand into energy bands which form as a result of the overlap between atomic orbitals. In this model, bands of energies are shown to change as the electrons' momentum across the material [24]. This energy difference is referred to the band gap of materials and is seen between the top of the valence band (VBM) and the bottom of the conduction band (CBM). Electronic and optical behavior of a material is strongly influenced by their band gap. Mathematically band gap Eg is calculated as:

$$E_g = E_{CBM} - E_{VBM} \qquad (1)$$

Electrically, the presence of a band gap allows you to excite electrons from the valence to the conduction band, making the band able to help with electric currents.

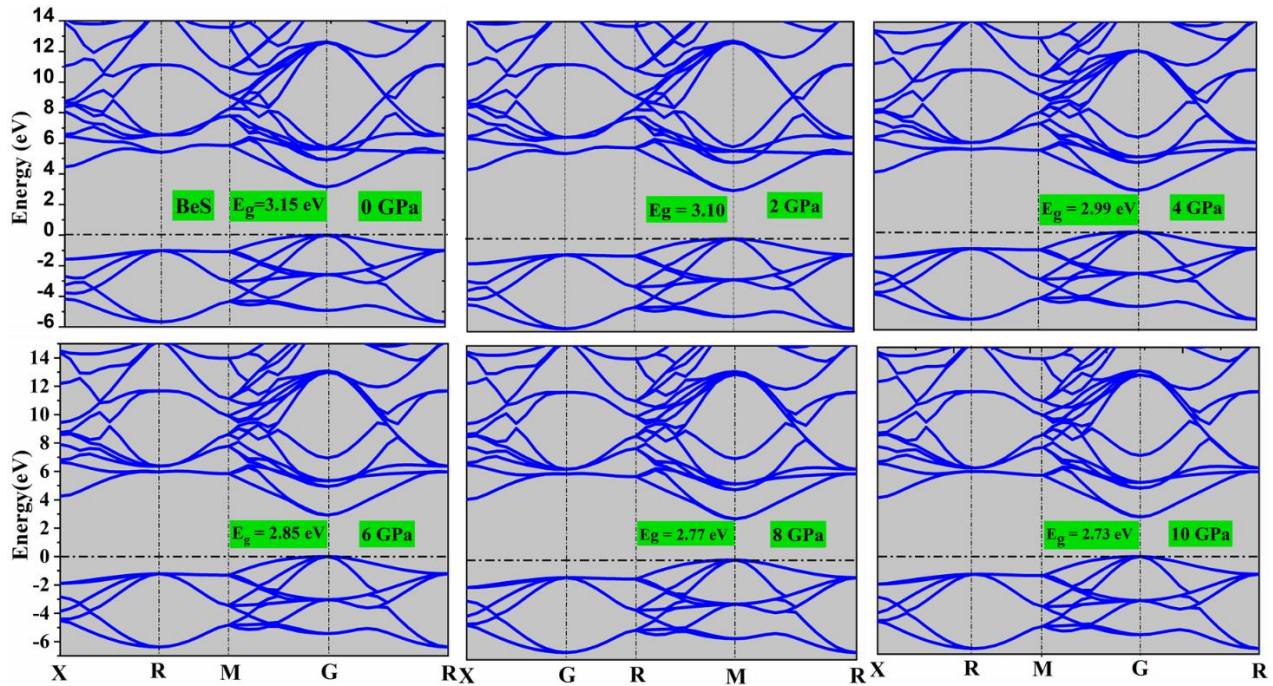

**Fig.7.** The band structure of BeS under varying pressure.

The electronic band structures of cubic BeS along the main Brillouin zone directions (X→R→M →G→R) are shown in Figure 7 for external pressures between 0 and 10 GPa. When pressure is applied, clear changes in the band dispersion and energy gap can be seen. These results suggest that the electronic structure of BeS changes significantly when under pressure such as a possible change in its band gap. Such actions demonstrate the material's propensity for changes when squeezed and its promise for modified electronic and optoelectronic features. The electronic band structure of BeS clearly divides the valence and conduction bands by the Fermi level fixed at 0 eV. The valence band goes from 0 to about -6 eV and the conduction band is from 0 to roughly 14 eV. The points with the highest energy in the valence band (VBM) and the lowest energy in the conduction band (CBM) are different in this pressure interval, located at canonical points of the Brillouin zone (G). The variation in band gap, we can confirm that BeS has a direct band gap at all pressures considered here.

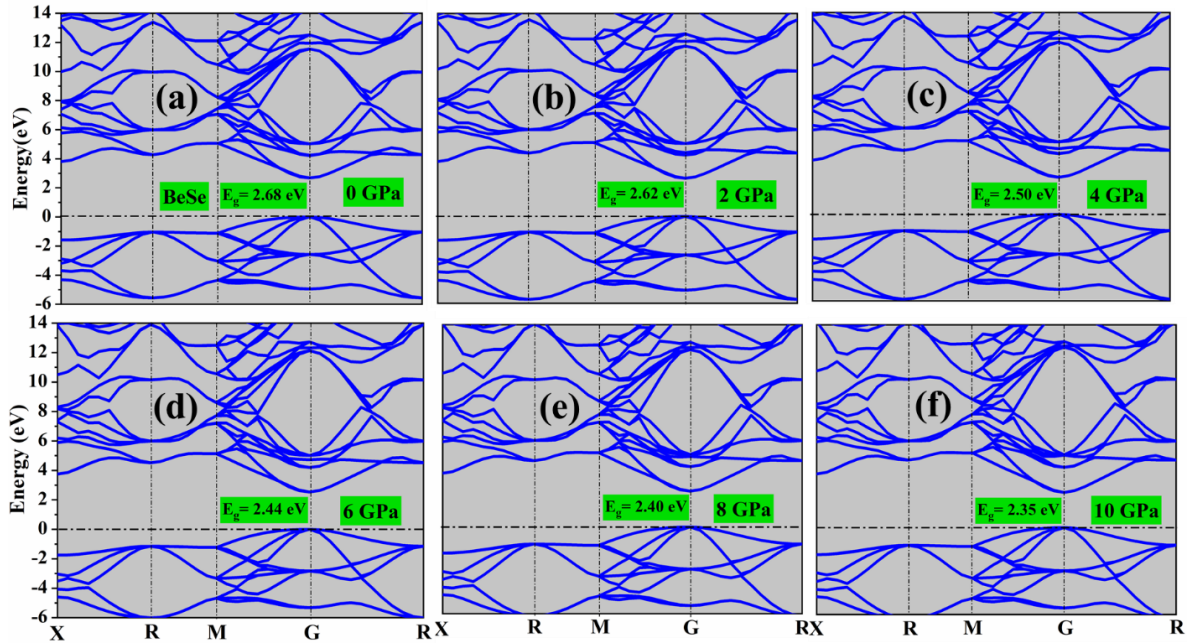

**Fig.8.** the electronic band structure of BeSe under varying pressure.

This observation suggest that the Bes have direct band gap and has semiconductor behavior when pressure range applied from 0 to 10 GPa. Similarly figure 8 shows electronic band structure of BeSe under the effect of different pressure. The calculated electronic band gaps for BeSe under pressures of 0, 4, 6, 8 and 10 GPa are 2.68, 2.62, 2.50, 2.44, 2.20 and 2.35 eV, respectively. The results shows that the band gap shrinks in proportion to the increase in applied pressure. Maximum compression in a semiconductor causes its bands to move closer together, indicating better mixing of atomic orbitals. As a consequence, the edge at which the material absorbs optical light shifts closer to the spectrum used for visible light, possibly making it valuable in pressure-controlled optical devices. However the same pattern was observed for BeTe as shown figure 9. The calculated band gap value for BeTe under pressure of 0, 4, 6, 8 and 10 GPa are 2.37, 2.33, 2.27, 2.2, 1.96 and 1.94 eV, respectively. Figure 10 (a-b) shows the band gap value and final energy under pressure 0-10 GPa as function of hydrostatic pressure. The final energy changes with applied pressure on BeX (X = S, Se, Te) is shown in figure 10 clearly by the graph. While raising the pressure from zero to 10 GPa, the total energy goes down, suggesting that higher pressure causes lower final energy. It appears that the crystal structure of BeX (X = S, Se, Te) becomes more energy efficient at higher pressure. Stability is best observed at 10 GPa about -2405.2357 eV total energy is the lowest.

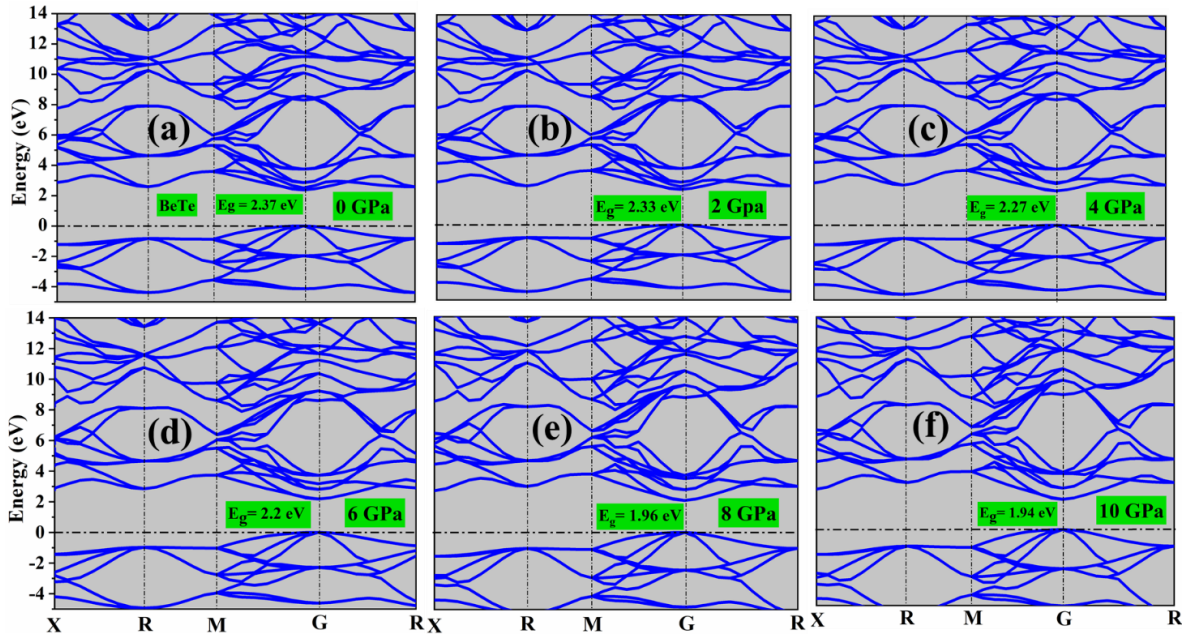

**Fig.9.** the electronic band structure of BeTe at multiple pressure.

### 3.2.2. Density of states (DOS)

The electronic structure of materials and its various spectral energy levels can be studied by the DOS [25]. It means the number of electronic states per energy level within the energy range can be presented. This data is significant due to how these states' distribution can change the electronic, optical and magnetic functionality of a material. DOS is widely considered to exist in two important forms. TDOS means total density of states which considering the densities of all available electronic states and partial density of States (PDOS) divides the total states into the portions created by various atoms or orbitals [26, 27].

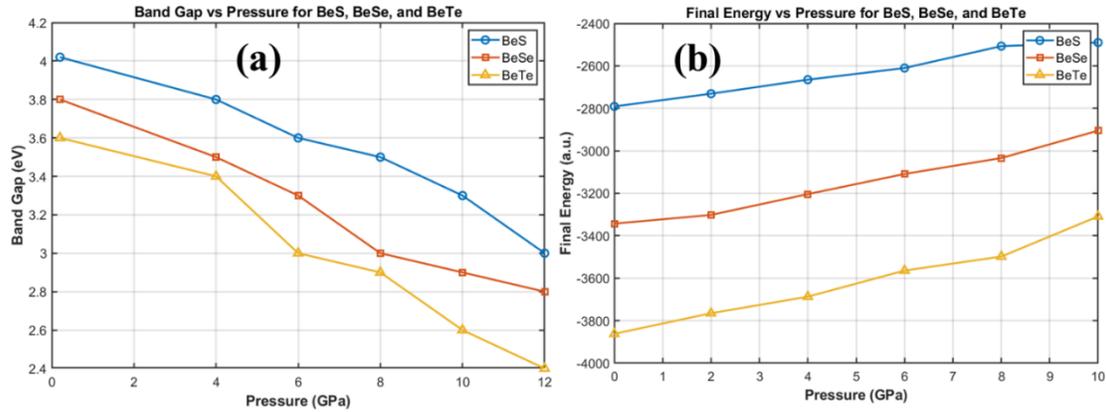

**Fig.10.** the calculated (a) band gap value and (b) final energy of BeX (X = S, Se, Te) at various pressure.

Total Density of States is projected for BeX (X = S, Se, Te) through the normal range at varying applied pressures, spanning an energy window of -4 eV to 14 eV as shown in figure 11. Fermi level which has a measured value of 0 eV, separates the VB from the CB effectively. The energy range for the valence band is from -4 eV to 0 eV and the conduction band starts from 0 eV to 14 eV. Importantly, the valence band is especially active near the intervals of -4 to 0 eV and -1 to 0 eV, pointing to the major effects of Be and the chalcogen elements (S, Se, Te) on the electronic structure. Pressure impacts the structure and population of electronic states, helping us see how this material binds and behaves electrically. The contribution by the valence band of BeX (X = S, Se, Te) compounds is most noticeable near -1 eV when ambient pressure is used. At 6 GPa in pressure, the dispersion of the VB drops, indicating that electronic states are moved differently. As the pressure reaches 6 GPa, the VB begins to add to the intensified results, peaking close to -2 eV at pressures just above 10 GPa. The trend demonstrates that the electronic structure of these metals shifts in a pressure-dependent way at lower energies.

However, according to the calculations, the CB still stays inert at the Fermi level at all levels of compression, confirming an insulating or semiconducting state. The most important CB contributions are regularly seen at 3-14 eV of energy. At zero GPa, the heat conduction follows a peak value at about 6 eV. When the pressure rises, CB contribution at this level drops rather than remaining sharp, making the peak less obvious. It shows that the electronic structure of BeX (X = S, Se, Te) compounds is influenced by external pressure, changing both their valence and conductance.

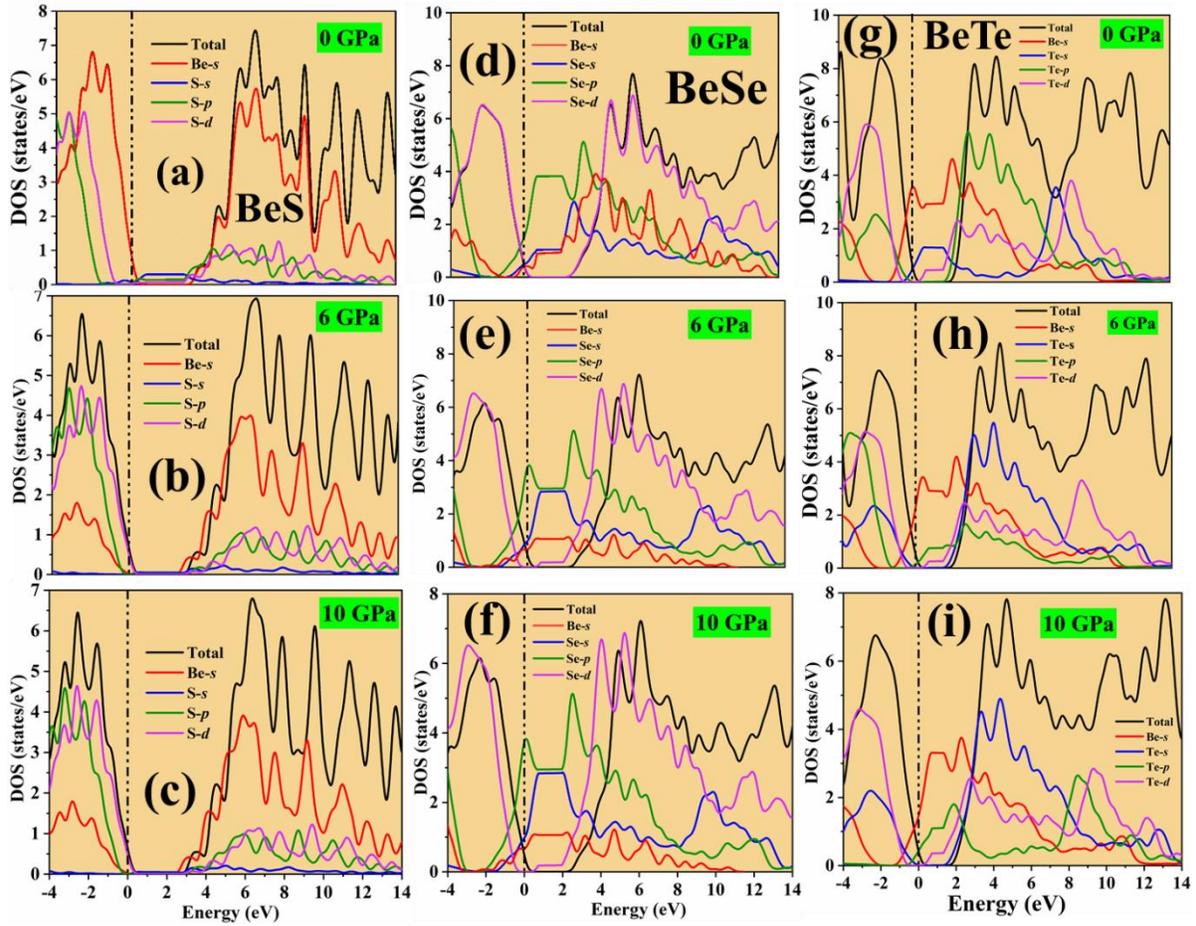

**Fig.11**. TDOS and PDOS of BeX (X = S, Se, Te) under applied pressure.

The value of the PDOS across an energy range of -4 to 14 eV for BeS are shown in figure 11 (a-c) at 0, 6 and 10 GPa of pressure. When pressure is 0 GPa, the Be-*s*, S-*p* and S-*d* orbitals (-2 to 0 eV) which are very prominent in the valence band (VB) region. The contributions to the CB are substantial from Be-*s*, S-*p* and S-*d* and orbitals and the combination of S-*s* orbitals which hints at mixing and hybridization in this high energy band. At 6 GPa, the behavior of BeS becomes noticeably involved, with their S-*p* and S-*d* orbitals near the Fermi level doing much of the work. It demonstrates that the orbitals begin to fit together much better as pressure causes as the distribution of states in the system is altered. In the crystal system, Be-*s*, S-*p* and S-*d* orbitals become more prominent, due to pressure increases their role in electronic transitions. However pressure at 10 GPa, both atomic and orbital components shrink, showing that a rearrangement of electron configuration or a shift in energy is possible. The Be-*s*, S-*p* and S-*d* orbitals continue to offer substantial contribution in the VB. CB prefers to place atoms in Be-*s*, S-*p* and S-*d* orbitals having more contribution and leaving S-*s* with not much impact.

Similarly PDOS for BeSe are shown in figure 11(d-f) at same energy range. The states in the valence band above negative values are dominated by the Se-*d* orbital, suggesting that both the strong localization and the bonding from Se-*p* and Se-*d* states play a major role. Trace amounts of Be-*s* and Se-*s* character have been noticed which shows that some of these states are taking part in the valence band structure. All orbitals contribute in the CB, lying over the Fermi level, since the electrons are found to be more dispersed and mixed. Nevertheless, the Se-*d* and Se-*p* orbitals are essential since they strongly influence the compound's conduction and optical transitions. In all three pressure points (0, 6 and 10 GPa), Se-*d* orbitals provide most of the valence band and both Se-*d* and Se-*p* are major contributors to the conduction band. Because of pressure, the intensities and energy positions of the BeSe orbitals change which leads to differences in both the electronics and optical properties of the compound.

In the same way PDOS for BeTe are shown in figure 11 (g-i) at same energy range. The Te-*d* and Te-*p* orbitals is significant in the valence band when pressure is 0 GPa. When pressure increases, the orbitals interact better which pushes their electrons closer and can result in wider and slightly changed bands. Be-*s* and Te-*s* orbitals keep their small secondary contributions and a little increase in orbital overlap could be caused by pressure effects. When pressure increases the Te-*d* orbital takes over as the main contributor to the valence band, making it clear its key role in bonding. Yet, boosting compression can shift the state densities a little which can move the peaks of the PDOS. Even though Be-*s* and Te-*s* orbitals add only gently, they always do so reliably. In the conduction band at 10 GPa, all orbitals play a role and the Te-*d* and Te-*p* orbitals continue to stand out the most. Te-*s* may be seen a little more clearly as the pressure increases, due to more influence from the orbitals of other elements.

### 3.3. Optical properties

The electronic structure of BeX (X = S, Se, Te) is the main part of its optical characteristics which deciding its response to light. Researchers describe this interaction with the complex dielectric function which goes by $\varepsilon(\omega)$, where $\omega$ is the frequency of the incoming photons. This dielectric function consists of two different sections real part $\varepsilon_1(\omega)$ analogous to the material's dispersion which means how the phase velocity of light changes as the frequency changes and the other is imaginary part ε2 (ω) which shows the ratio of energy lost by the material as it absorbs the energy. By using Kramer's and Kronig relations [28]. Mathematically it can be expressed as:

$$\epsilon(\omega) = \epsilon_1(\omega) + i\epsilon_2(\omega) \qquad (2)$$

Generally, dielectric response in electronic materials results from either inter-band or intra-band electron transitions. Within metals, transitions taking place within a single energy band happen most frequently. During inter-band transitions, electrons move from one energy band to another and may do so directly, leaving the momentum equal or indirectly, when they acquire or lose some momentum by working with phonons. The $\varepsilon_2$ (ω) of the dielectric function describes the collective behavior of electrons moving among filled and empty states in a material [23]. In a different way, the real part, $\varepsilon_1$ (ω), handles the electronic polarization inside the material and can be found by applying the Kramers-Kronig dispersion relation. The behavior of ε2 (ω) is explained by analyzing dipole transition matrix elements involving electron excitations [29, 30].

$$\varepsilon_1(\omega) = 1 + \frac{2}{\pi}\int_0^\infty \frac{\varepsilon_2(\omega)\omega' d\omega'}{\omega'^2 - \omega^2} \qquad (3)$$

$$\varepsilon_2(\omega) = \frac{Ve^2}{2\pi m^2 \omega^2}\int d^3k \sum_{n,n'} \langle k,n|p|k,n'\rangle^2 f(kn)[1-f(kn')]\,\delta(E_{kn} - E_{kn'} - \omega) \qquad (4)$$

In these terms, Ei is the energy of an electron and M refers to a dipole matrix element responsible for the chances of optical transitions. In a transition, i and j serve as symbols for the electronic state at the beginning and at the end. Furthermore, fi is used to show the probability of an initial state being occupied according to the Fermi-Dirac distribution at a particular temperature. The total impact of each of these parts is seen in the dielectric function, describing a material's optical performance.

The EM or light waves travel through a material can be measured by using a refractive index called n (ω) [28].

$$n(\omega) = \left[\frac{\sqrt{\varepsilon_1^2(\omega)+\varepsilon_2^2(\omega)}}{2} + \frac{\varepsilon_1(\omega)}{2}\right]^{1/2} \qquad (5)$$

the complex dielectric function, gives other quantities such as absorption, energy loss function, refractive index, extinction coefficient, real part of optical conductivity and reflectivity.

$$k(\omega) = \left[\frac{\sqrt{\varepsilon_1^2(\omega)+\varepsilon_2^2(\omega)}}{2} - \frac{\varepsilon_1(\omega)}{2}\right]^{1/2} \quad (6)$$

$$L(\omega) = -Im\left(\frac{1}{\varepsilon(\omega)}\right) = \frac{\varepsilon_2(\omega)}{\varepsilon_1^2(\omega)+\varepsilon_2^2(\omega)} \quad (7)$$

$$R(\omega) = \left|\frac{\tilde{n}(\omega)-1}{\tilde{n}(\omega)+1}\right| = \frac{(1+n)^2+k^2}{(1-n)^2+k^2} \quad (8)$$

$$\alpha(\omega) = \sqrt{2\omega}\left[\sqrt{\varepsilon_1^2(\omega)+\varepsilon_2^2(\omega)} - \varepsilon_1(\omega)\right]^{1/2} \quad (9)$$

The above equations provide extinction coefficeint, energy loss, reflectivity and absorption of materials respectively.

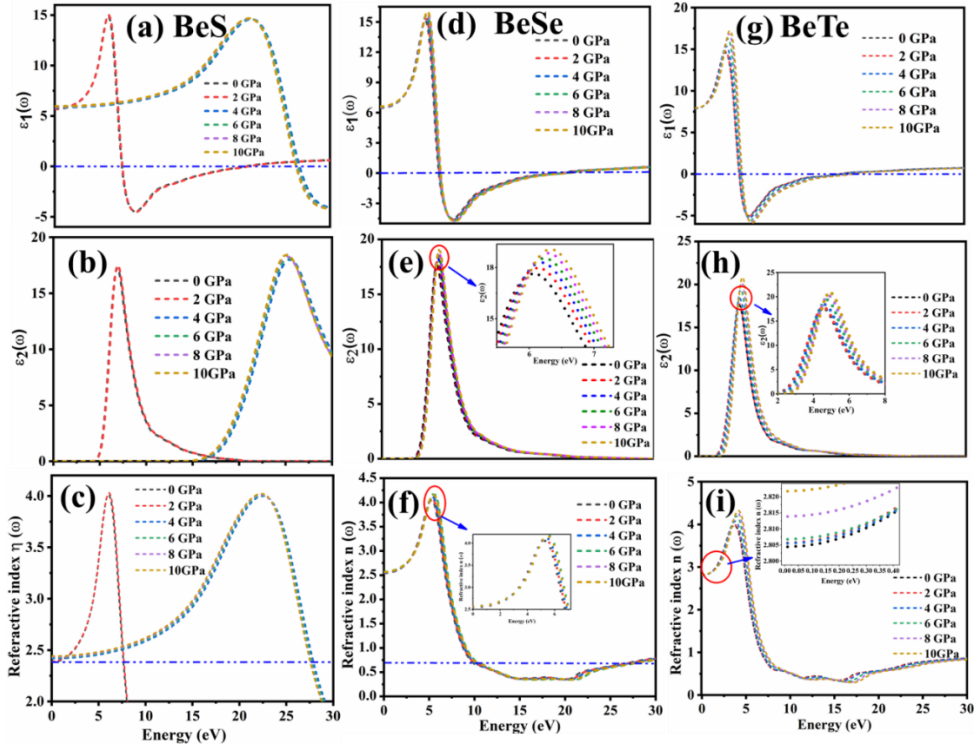

**Fig.12.** the computed dielectric function $\epsilon(\omega)$ and refractive index $n(\omega)$ of (a-c) BeS (d-e) BeSe and (g-i) BeTe at different pressure.

Figure 12 (a-c) shows real and imaginary part of dielectric function of BeS with spectra of refractive index n ($\omega$) at various pressure 0-10 GPa. According to calculations, the BeS structure shows that at zero frequency, the static dielectric constant $\varepsilon_1(0)$ is not zero. The $\varepsilon_1(0)$ values are

found 5.1658, 5.1845, 5.2585, 5.2813, 5.2721 and 5.27443, which are corresponding to applied pressures from 0 to 10 GPa in steps of 2 GPa as shown in figure 12(a). The highest peak values for BeS are observed at 4.0051, 4.0263, 14.599, 14.6308, 14.6502 and 14.6921 eV under applied pressure of 0, 2, 4, 6, 8 and 10 GPa respectively. The results shows that the largest polarization happens at the given point and pressures. When energy steps up into the higher ultraviolet (UV) range, polarization lessens and starts to display negative values at 7.2840, 7.3734, 21.4132, 21.4521, 21.5049 and 21.518 eV for employed pressure from 0 to 10 GPa. The shift of the peak value to higher energy in the real part of the dielectric function $\varepsilon_1(\omega)$ for BeS under increasing pressure is primarily due to the pressure-induced modifications in the electronic band structure of the material.

The imaginary part of dielectric function $\varepsilon_2(\omega)$ are shown in figure 12(b). The optical absorption spectra are found at energies of 4.753, 4.734, 16.2134, 16.1023, 16.1242 and 16.0226 eV for pressures between 0 and 10 GPa with each increase being by 2 GPa. By using GGA+PBE, the strongest peaks appear at 7.1103, 7.2064, 25.2663, 25.1171, 25.0061 and 25.1382 eV at pressures of 0, 2, 4, 6, 8 and 10 GPa, respectively. Most of these peaks are produced as the system experiences changes between the S-*p* and S-*d* electronic states. It is observed from the spectra that UV radiation causes the strongest photon absorption. In figure 12(c), the behavior of refractive index n($\omega$) dispersion can be seen. The temperature sensitivity n($\omega$) of BeS at pressures from 0 up to 10 GPa. The spectra of the dynamic value of n($\omega$) is computed from the known static index values of the material. When 0, 2, 4, 6, 8 and 10 GPa pressure are applied the static refractive index n($\omega$) values for n(0) are 2.3832, 2.3913, 2.4171, 2.4255, 2.4350 and 2.4426, respectively. With increased of photon energy. At the set pressure, the energy increases n($\omega$) from its usual value and rises strongly at 6.14911, 6.0601, 22.09, 22.1312, 22.1243 and 22.1132 eV.

Similarly, the dielectric function and refractive index of BeSe and BeTe are shown in figure 12(d-f) and figure 12(g-i) respectively. the BeSe structure shows that at zero frequency, the static dielectric constant $\varepsilon_1(0)$ is not zero. The $\varepsilon_1(0)$ values are found 6.4854, 6.4949, 6.5192, 6.5859, 6.5342 and 6.6163, which are corresponding to applied pressures from 0 to 10 GPa in steps of 2 GPa. in the same way The $\varepsilon_1(0)$ values for BeTe 6.2432, 6.2156, 6.2134, 6.2124 and 6.2173, 6.2376 at elevated pressure 0 to 10 GPa respctively. The highest peak values for BeSe are observed at 5.0382, 5.0512, 5.1545, 5.2193, 5.2283 and 5.2711 eV and for BeTe the highest values

are 5.1352, 5.1452, 5.1521, 5.2132, 5.2243 and 5.2643 eV under different pressure 0 to 10 GPa. the negative values for both compound confirm that energy are steps up into the higher ultraviolet (UV) range. The imaginary part of dielectric function $\varepsilon_2(\omega)$ and refractive index n ($\omega$) for BeSe and BeTe are shown in figure 12(e and f) and figure (h and i) respectively. The optical absorption spectra for BeSe are found at energies of 5.4942, 5.5153, 5.5526, 5.5900, 5.6149 and 5.6523 eV. Similarly absorption spectra for BeTe are 3.1432, 3.1265, 3.1375, 3.13865, 3.1745 and 3.1683 eV at pressures between 0 and 10 GPa with each increase being by 2 GPa. The strongest peaks appear at 6.1504, 6.2251, 6.2002, 6.2998, 6.2625 and 6.3870 eV for BeSe and highest peak values for BeTe are 4.1432, 4.2543, 4.4531, 4.2543, 4.5234 and 4.2532 at pressures of 0, 2, 4, 6, 8 and 10 GPa, respectively. It is observed from the spectra that UV radiation causes the strongest photon absorption in both compounds. In figure 12(f and i), the behavior of refractive index n($\omega$) dispersion for BeSe and BeTe can be seen. The spectra of the dynamic value of n($\omega$) is computed from the known static index values of the material. When 0, 2, 4, 6, 8 and 10 GPa pressure are applied the static refractive index n($\omega$) values for BeSe n(0) are 2.5432, 2.5486, 2.5532, 2.5654, 2.55723 and 2.4426, respectively with photon energy. At the set pressure, the energy increases n($\omega$) for BeSe from its usual value rises strongly at 5.0466, 5.0558, 5.1564, 5.1236, 5.1054 and 5.2745 eV. Similarly strong peaks for BeTe are 4.2216, 4.0135, 4.1264, 4.1492, 4.1362 and 4.2741 eV as shown in figure 12 (i).

Figure 13 (a-f) displays the reflectivity R($\omega$) and energy loss L($\omega$) spectra for BeS, BeSe and BeTe compounds. The value of R($\omega$) was measured at pressures from 0 to 10 GPa, which assisting to understand the surface qualities of the material as stated in reference [48]. The static reflectivity values that were computed for BeS reported R(0) as 0.1623, 0.1134, 0.11719, 0.1732, 0.1745 and 0.1758 for pressures of 0, 2, 4, 6, 8 and 10 GPa, respectively. Significant reflectivity peaks at

5.3057, 5.4152, 25.3673, 25.4308, 25.49677 and 25.4597 eV are observed in the spectra for each pressure increase in order. The graph shows significant peaks difference for initial and final applied pressure. Each peak displays an energy reflection level and these different peaks are a result of anisotropy of the material when pressure is applied from different sides. Similarly the reflectivity spectra for BeSe and BeTe are shown figure 13 (c &e). the static reflectivity R(0) for BeSe 0.1900, 0.1909, 0.1916, 0.1924, 0.1933, and 0.194 at applied pressure of 0 to 10 Gpa respectively. the peak values are correspond to 9.2550, 9.2084, 9.1851, 9.2783, 9.2543 and 9.2465 eV for photon energy under the applied pressure. In figure 13(e) the static reflectivity R(0) for BeTe is recorded as 0.2249, 0.2250, 0.2251, 0.2252, 0.2262 and 0.2272 while the corresponding values are 7.2571, 7.2640.7.2898, 7.2756, 7.2876 and 7.2943eV under applied applied pressure.

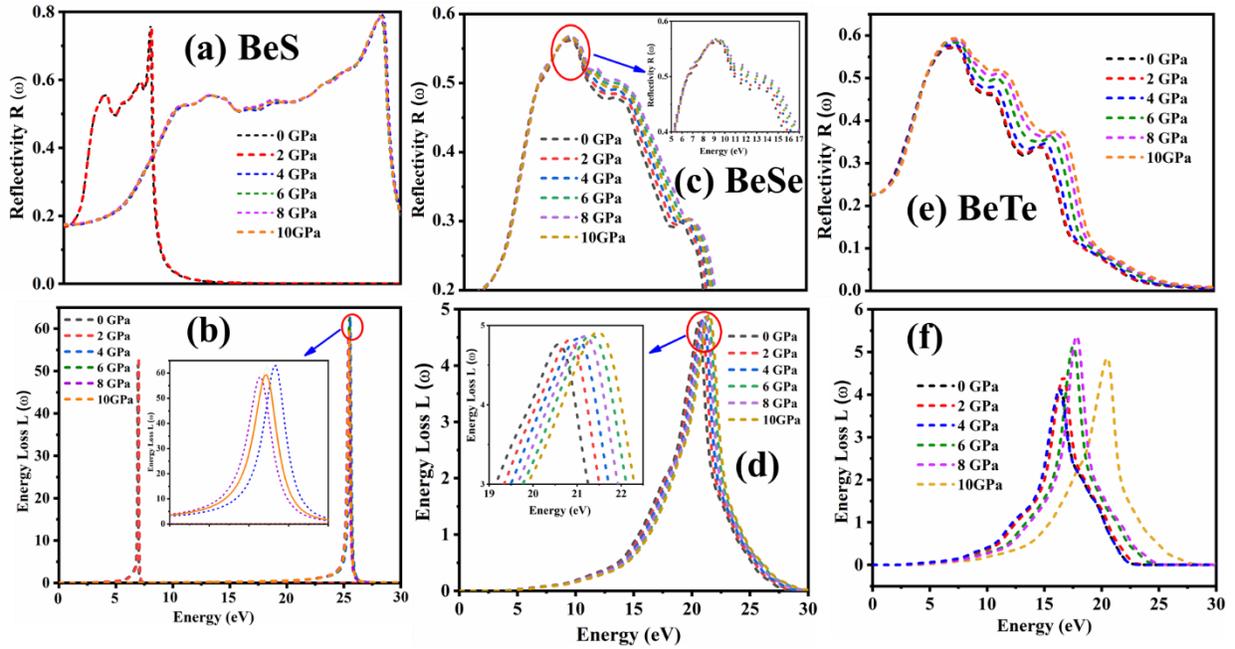

**Fig.13.** the reflectivity R($\omega$) and energy loss L($\omega$) of (a-b) BeS (c-d) BeSe and (e-f) BeTe under different pressure.

The plasma frequency in a material can be determined through energy loss L($\omega$) which depends on electron energy. It gives useful insights into the manner in which fast electrons gradually lose energy in the system. When photons have strong enough energy they collide with the material so electrons are not restrained in the crystal lattice anymore. Instead of this the energy loss starts to vary in a special rhythm and the peaks seen on the spectral curve are related to the plasmons the oscillations in these collective modes. The figure 13 (b,d and f) shows energy loss L(ω) function for BeS, BeSe and BeTe compounds. At pressures from 0 to 10 GPa, the spectra are observed for BeS at energy values starting at 5.3421, 5.2343, 25.1635, 25.1039, 25.0422 and 25.2113 eV at pressures of 0, 2, 4, 6, 8 and 10 GPa, respectively. The maximum peaks in the spectra show up at 5.4821, 5.4243, 26.1430, 26.1513, 16.1990 and 26.2458 eV. According to this trend, energy loss is highest in the portion of the ultraviolet (UV) spectrum from 24 to 25.3 eV. Similarly energy loss function for BeSe shows initial energy values of 6.4284, 6.4224, 6.4221, 6.4234, 6.4242 and 6.4242 at applied pressure of 0, 2, 4, 6, 8 and 10 GPa respectively. The peak values are observed at 20.5178, 20.7527, 20.9171, 21.1050, 21.2694 and 21.4808 eV at given pressure. Based on this trend, energy loss occur within the UV range between 19 to 22 eV. The energy loss function for BeTe also has initial energies of 6.4514, 6.4523, 6.4543, 6.4552, 6.4542 and 6.4523 eV at applied pressure. When increasing pressure to 2, 4, 6, 8 and 10 GPa the peak values are at 16.3953, 16.4842, 16.3064, 17.3734, 17.8081 and 20.4856 eV. The corresponding peak positions are found that BeTe has most of its energy lost in the ultraviolet area, mainly between energy values of 10 and 27 eV.

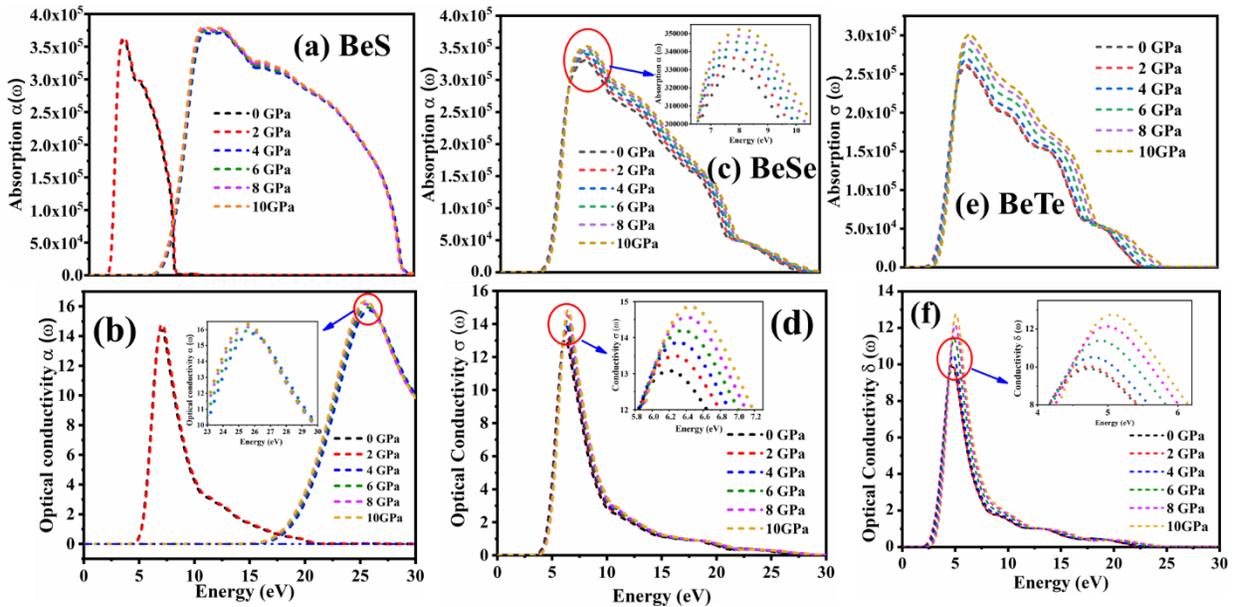

**Fig.14.** the absorption α(ω) and optical conductivity σ(ω) of (a-b) BeS (c-d) BeSe and (e-f) BeTe against under the effect of different pressure.

The absorption coefficient is shown by α(ω), denotes the degree to which a material absorbs light or electromagnetic radiation. A high value means the material can take in more different types of light. Essentially, It shows us the efficiency with which a material absorbs light and gives off radiation in other forms, with the efficiency feeling the frequency (or energy) of incoming light radiation [31].in figure 14(a-f) shows the absorption and optical conductivity spectra of BeS, BeSe and BeTe compounds. The absorption spectra of BeS initially shows absorption starting at 4.9247, 4.8563, 5.3454, 5.3702, 5.2632 and 5.29023 eV with pressure ranging from 0 to 2 GPa respectively. it is reported that $\alpha(\omega)$, has peak values of 2.5342 ($\lambda$=490 nm) and 2.6432 ($\lambda$=469 nm) eV for 0 and 2 GPa respectively. As the pressure increases from 4 to 10 GPa the peak values shift toward high energies with value of 5.4231 ($\lambda$=228 nm), 5.43612 ($\lambda$=228 nm), 5.4164 (229 nm) and 5.4532 eV (227 nm). The wavelength values at 0 and 2 GPa indicate that most of absorption occur in visible region, while for pressure ranging from 4 to 10 GPa the absorption shifts to the UV region which is favourable for optoelectronic applications.

Similarly, the absorption spectra of BeSe begin at energy values of 3.6183, 3.6245, 3.62431, 3.61523, 3.64234 and 3.6534 eV under elevated pressure ranging from 0 to 10 GPa. the peak values are observed ate 7.7485 ($\lambda$=160 nm), 7.7817 ($\lambda$= 159 nm), 7.8481 ($\lambda$= 158 nm), 7.8813 ($\lambda$= (157 nm), 7.9145 ($\lambda$= 156 nm) and 7.9478 eV ($\lambda$= 156 nm). these values indicate that absorption occur within the UV region. In contrast, the absorption spectra of BeTe shows initial energy values of 2.0519, 2.1243, 2.1472, 2.3634, 2.15801 and 2.4203 eV when subjected to elevated pressure between 0 to 10 GPa. The observed peak values are observed ate 5.8207 ($\lambda$=213 nm), 5.9097 ($\lambda$= 209 nm), 6.0875 ($\lambda$= 203 nm), 6.1764 ($\lambda$= 200 nm), 6.2652 ($\lambda$= 197 nm) and 6.4432 eV ($\lambda$= 192 nm). these energy levels suggest that absorption primarily occur in the UV region

In conclusions all of these elements exhibit strong absorption of UV light, with BeS switching from visible light to UV absorption as pressure increases, while BeSe and BeTe being even stronger in the UV region. As results, all these materials are highly suitable for optoelectronics devices such as UV sensors or lasers and other UV based devices.

The function $\sigma(\omega)$, called the optical conductivity, tells us how a material conducts electricity in response to electromagnetic (EM) radiation. It explains the reaction of the material to light of several frequencies and uncovers information about how charge carriers such as electrons, flow with that radiation. The data in figure 14 (b,d,f) shows the optical conductivity of BeX (X = S, Se, Te). The $\sigma(\omega)$ behavior of BeSe shown in figure 14 (b) when subjected to pressures applied from 0 to 10 GPa. At each pressure 0, 2, 4, 6, 8 and 10 GPa conductivity starts to grow when the

materials have been pushed at the thresholds of 4.3775, 4.9110, , 16.8878, 17.6575, 17.8732 and 17.9069 eV. The BeS exhibit optical conductivity at energy value of , 7.0861, 7.1876, 25.5882, 25.4993, 25.3214 and 25.2326. The results indicate that BeSe has more optical conductivity in the ultraviolet region than in the visible part of the spectrum. Similarly In figure 14(d), the optical conductivity of BeSe is shown. The σ(ω) of a BeSe at higher pressures, the appearance of conductivity occurs at higher energies and showing threshold values measured at 3.2765, 3.2210, 3.2487, 3.2543, 3.2567 and 3.2906 eV. The conductivity increases in the ultraviolet area and remains nearly absent in the visible section. The BeTe also demonstrates unique optical conductivity under pressure as shown in figure 14 (f) and there are peaks detected at 2.1067, 2.1025, 2.2104, 2.2764, 2.2673 and 2.2875 eV. The strong UV response confirmed by these values makes both materials suitable for use in UV-based optoelectronics.

**Table.1.** the calculated parameters of BeX (X = S, Se, Te) compared with previous work.

| Compound | Pressure (GPa) | Lattice constant $a_o$ (Å) | Band gap ($E_g$) | Static $\varepsilon_1(0)$ | Static n (0) | Static R (0) | Symmetry | Approach | Reference |
|---|---|---|---|---|---|---|---|---|---|
| BeS | 0 | 4.0 | 3.15 | 5.1658 | 2.3832 | 0.1623 | Cubic F-43m | CASTEP | This work |
| | 2 | 3.8 | 3.1 | 5.1845 | 2.3913 | 0.1134 | | | |
| | 4 | 3.6 | 2.99 | 5.2585 | 2.4171 | 0.1171 | | | |
| | 6 | 3.5 | 2.85 | 5.2813 | 2.4255 | 0.1732 | | | |
| | 8 | 3.23 | 2.77 | 5.2721 | 2.4350 | 0.1745 | | | |
| | 10 | 3.0 | 2.73 | 5.2744 | 2.4426 | 0.1758 | | | |
| BeSe | 0 | 3.8 | 2.68 | 6.4854 | 2.5432 | 0.1900 | Cubic F-43m | CASTEP | This work |
| | 2 | 3.5 | 2.62 | 6.4949 | 2.5486 | 0.1909 | | | |
| | 4 | 3.3 | 2.52 | 6.5192 | 2.5532 | 0.1916 | | | |
| | 6 | 3.0 | 2.44 | 6.5859 | 2.5654 | 0.1924 | | | |
| | 8 | 2.9 | 2.40 | 6.6163 | 2.5572 | 0.1933 | | | |
| | 10 | 2.8 | 1.96 | 6.6130 | 2.4426 | 0.1945 | | | |
| BeTe | 0 | 3.6 | 2.37 | 6.2432 | 2.5743 | 0.2249 | Cubic F-43m | CASTEP | This work |
| | 2 | 3.4 | 2.33 | 6.2156 | 2.5867 | 0.2250 | | | |
| | 4 | 3.0 | 2.27 | 6.2134 | 2.5779 | 0.2251 | | | |
| | 6 | 2.9 | 2.22 | 6.2124 | 2.5743 | 0.2252 | | | |
| | 8 | 2.6 | 1.96 | 6.2173 | 2.5765 | 0.2262 | | | |
| | 10 | 2.4 | 1.94 | 6.2376 | 2.5784 | 0.2272 | | | |
| BeS | - | 3.499 | 4.93 | 4.68 | 2.17 | - | | | |
| BeSe | - | 3.775 | 3.74 | 6.91 | 2.53 | - | Tetragonal | WIEN2K | [32, 33] |
| BeTe | - | 3.687 | 3.48 | 5.20 | 2.30 | - | | | |
| BeS | - | 4.870 | - | - | - | - | | | |
| BeSe | - | 5.137 | - | - | - | - | - | Exp | [2] |
| BeTe | - | 5.617 | - | - | - | - | | | |
| BeS | - | 4.800 | 2.78 | - | - | - | | | |
| BeSe | - | 5.085 | 2.23 | - | - | - | Hexagonal | LDA | [34] |
| BeTe | - | 5.557 | 1.47 | - | - | - | | | |
| BeS | | 4.806 | 5.46 | - | - | - | | | |
| BeSe | 69 | 3.280 | 4.44 | - | - | - | Cubic | LDA | [35, 36] |
| BeTe | | 5.563 | 3.59 | - | - | - | | | |
| BeS | - | 3.647 | 4.40 | - | - | - | | | |
| BeSe | - | 3.342 | 4.0 | - | - | - | Cubic | TB-mBj | [37] |
| BeTe | - | 5.412 | 2.40 | - | - | - | | | |

## 3.4. Thermodynamic properties

Thermodynamic properties of a materials are essential for understanding their thermal behavior under varying different conditions like pressure and temperature. The Gibbs free energy and Debye temperature of BeS, BeSe and BeTe compounds are shown in figures 15 (a-f). Thermodynamic properties are considered by analyzing them as temperature is changed for each of the given pressures, 0, 2, 4, 6, 8 and 10 GPa. Gibbs free energy which depends on both temperature and pressure, analogous to the phase stability of a material in thermodynamics. When the temperature is higher, it is seen that the differences in Gibbs free energy between various phases generally reduce under the influence of higher pressure. It means that high temperatures tend to lower the effect of pressure on the stability of phases which can result in phase transitions or groups of phases having similar energy values.

As we know,

$$G = H - TS \quad (10)$$

$$H = E + PV \quad (11)$$

From the above equations we get,

$$G = E + PV - TS \quad (12)$$

Take differentiation of (12),

$$dG = dE + PdV + VdP - TdS - SdT \quad (13)$$

We know,

$$dE + PdV = dH \quad \text{and} \quad TdS = dH \quad (14)$$

So, eq (13) becomes,

$$dG = dH + VdP - dH - SdT \quad (15)$$

Hence,

$$dG = VdP - SdT \quad (16)$$

Equation 16 is a relation in thermodynamics that shows how the pressure of a system adjusts as its temperature increases (for example, during the transition from solid to liquid or liquid to gas). Gibbs free energy allows for calculating latent heat and understanding different phases in a substance.

$$\left.\frac{dG}{dT}\right|_p = -S \quad \text{as} \quad -\left.\frac{dG}{dT}\right|_p = S \quad (17)$$

Since entropy always has a positive value, equation 17 means that Gibbs free energy decreases as temperature goes up. Therefore, as the temperature goes up, the Gibbs free energy becomes less positive (or more negative), indicating that a system tends to become less organized and more stable. The Gibbs free energy for BeS, BeSe and BeTe decreases with temperature which can be seen in figure.15 (a,c and e). Gibbs energy (G) will always decreases as the temperature (T) goes up. Because of this trend, it's clear that all three compounds BeS, BeSe and BeTe reach greater stability when the temperature rises. The thermodynamic relationship states ($\Delta H = \Delta G + T\Delta S$) The heat change ($\Delta H$) is the same as the change in Gibbs free energy. The Gibbs free energy ($\Delta G$) plus the result of temperature (T) and the entropy change points out that both energy and disorder work together to impact the stability of phases and the behavior of a material. It is noted for BeS, BeSe and BeTe that the enthalpy changes with pressure, with the highest and lowest values appearing at 0 and 10 GPa, respectively. Therefore, with increased pressure, these materials become more stable energetically which is revealed by a decline in enthalpy. The trend means that under higher pressure, the atoms interact more strongly and it becomes easier for them to bond energetically.

In figure.15 (b,d and f)), the impact of constant pressure 0, 2, 4, 6, 8 and 10 GPa on the Debye temperature ($\Theta_D$) of BeS, BeSe and BeTe is shown. The Debye temperature ($\Theta_D$) represents the highest possible frequency of motion (called a phonon mode) within a crystal lattice in thermodynamics. It helps to tie how a material bends (elastically) to its change in temperature properties. More specifically $\Theta_D$ ties in to thermodynamic aspects that consist of phonon actions, the area of thermal expansion, thermal conductivity, Gibbs free energy and lattice enthalpy. The Debye temperature decreases as the temperature increases which means that elevated heat causes the lattice vibrations to become softer. In addition, by relying on models depending on elastic constants and density, it is possible to predict the mean elastic wave velocity as well as the Debye temperature.

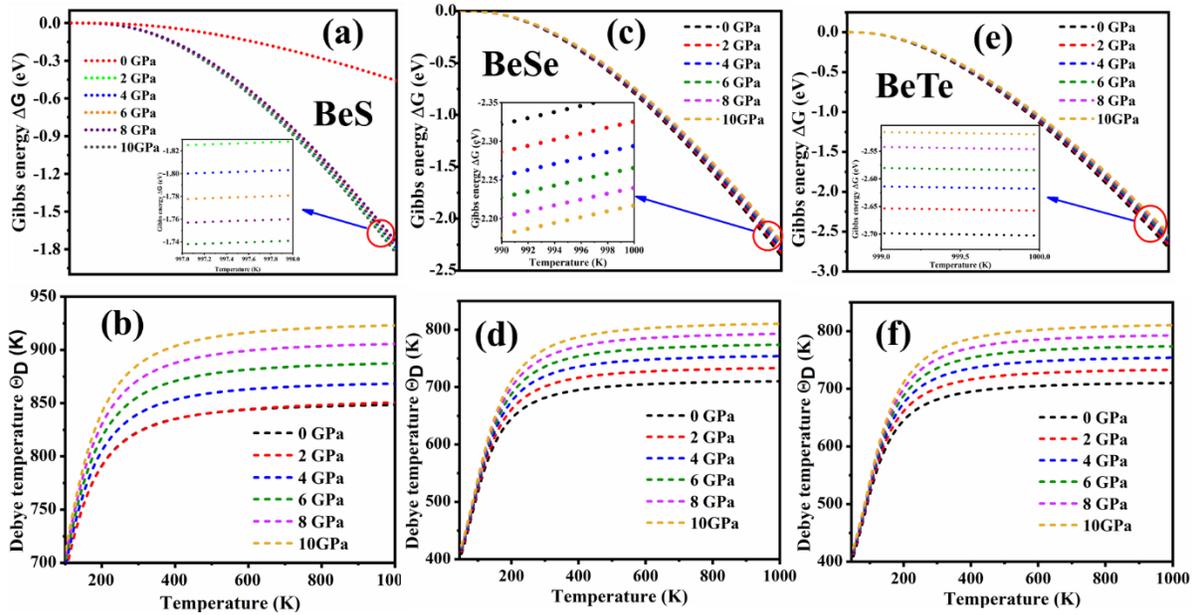

**Fig.15**. Illustration of Gibbs energy and Debye temperature of (a-b) BeS (c-d) BeSe and (e-f) BeTe under the effect of different pressures.

The rise in Debye temperature $\Theta_D$ for BeS, BeSe and BeTe when pressure are increases, it suggests that the crystal lattices become harder under pressure. When more pressure is applied, atoms come closer which increases the short-range opposing forces and the strength of the chemical bonding. Because of this, the sound velocity in the material goes up as the elastic moduli (bulk and shear) are usually greater at higher pressures. Because $\Theta_D$ depends on atom mass and elastic wave speed, the growth of chemical bonds and vibrations at higher pressure makes the Debye temperature grow. Due to this trend which boosts their strength and stability to pressures, these materials may be fit for use in thermal and mechanical situations that withstand pressure. In case of BeS, BeSe and BeTe: BeS (with lower atomic weights and stronger bonds) generally shows the most pronounced differences in $\Theta_D$ and fastest changes under pressure while BeSe and BeTe, being heavier, still see an increase in $\Theta_D$ under pressure, mostly due to bonding effects, but the boost is generally less significant than in BeS because they have a higher atomic mass and are more easily polarized.

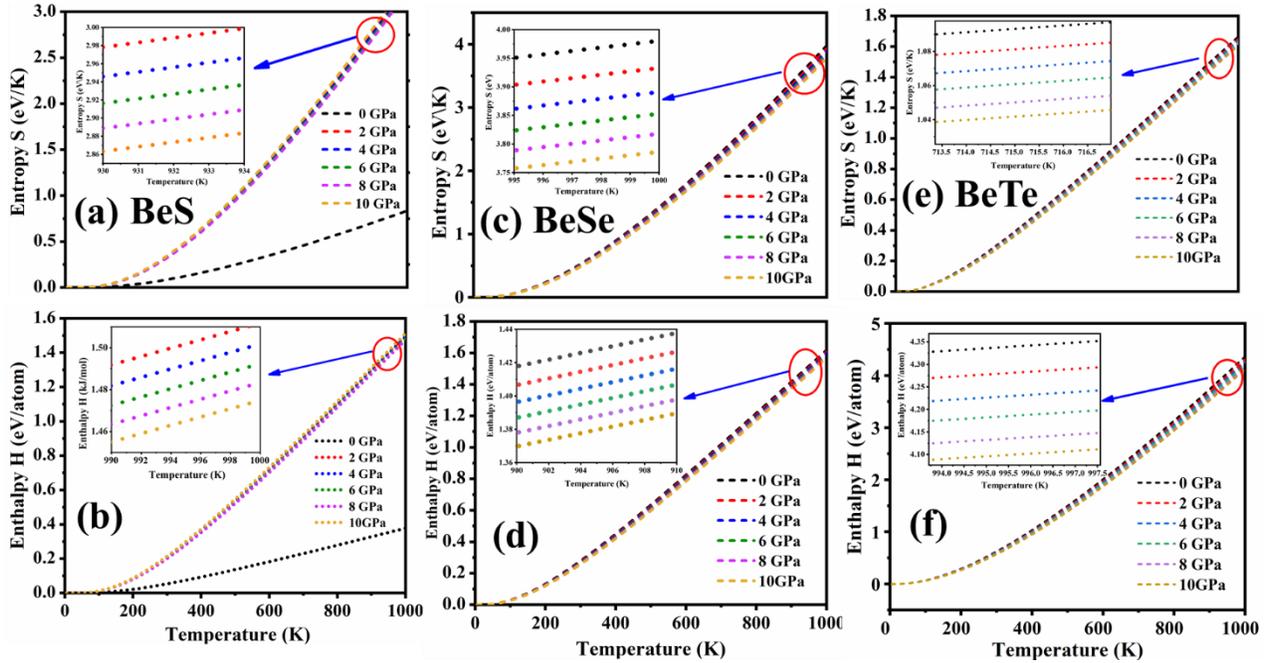

**Fig.16**. Illustration of entropy and enthalpy of (a-b) BeS (c-d) BeSe and (e-f) BeTe under the effect of different pressures.

The Helmholtz free energy (F) is used to calculate entropy (S) with the help of thermodynamic relation $S = -\left(\frac{\partial F}{\partial T}\right)_V$. When the volume is maintained as constant and P is zero, the change may be called evaluation at a constant volume and zero pressure (P = 0). The observed growth in entropy at higher temperatures for BeS, BeSe and BeTe as shown in figure.16 (a,c,and e) agrees with prospects since it is caused by increased motion of atoms in the system. When looked at from 0 to 100 K, all three materials show entropy going up which reflects increased activity and heat from phonons. As we know,

$$\frac{Q}{T} = S \quad \text{and} \quad T\,dS = dQ \quad \text{and} \quad dH = TdS \quad (18)$$

Q stands for the heat added inside the system. T is the absolute temperature. $S$ The quantity called entropy. Under stress, the amount of heat in the sample will change ($\Delta Q$) is the same as the change in enthalpy $\Delta H$.

$$H = \int_0^T C_p\, dT \quad (19)$$

So entropy,

$$S = \int_0^T \frac{C_p\, dT}{T} \quad as, \quad ds = \frac{C_p\, dT}{T} \quad (20)$$

$$\frac{ds}{dT} = \frac{C_p}{T} \quad (21)$$

Entropy (S) always increases with temperature (T) an important principle in thermodynamics though the speed of the increase may vary is lower the temperature. In this case, dS/dT indicates that the slope increases at a slow rate, the entropy increases along with temperature.

When the pressure rises (0 to 10 GPa), the overall entropy at a certain temperature goes down. This happens because increased pressure shrinks the atomic lattice, stops the atoms from vibrating freely and cuts back on the number of available arrangements. Therefore, the system becomes more organized and its level of entropy drops. Entropy reduction with stronger pressure can be observed in BeS, BeSe and BeTe. The extent of this happens differently depending on how light or heavy the element is and how strong its bonds are. The reason for lower entropy at 0 GPa in BeS is due to its stiff rigidity and high-temperature phonons which reduce the number of available states for phonons. Increasing the pressure by just a little bit softens the material which augments the amount of entropy. However, BeSe and BeTe are already less rigid at 0 GPa which means that pressure affects their entropy only a little.

The changes in pressure (P) and enthalpy (H are describe by following equation.

$$(\partial H/\partial P)_T = V \qquad (22)$$

So, when pressure is raised without adding heat, PV can still increase the enthalpy of the system. Here $(\partial H/\partial P)_T$ represents the partial derivative of enthalpy with respect to pressure at constant temperature (T), and it is equal to the molar volume (V). According to the Gibbs free energy relation, enthalpy is also related to entropy and free energy through the equation: ($\Delta H = \Delta G + T\Delta S$). in figure.16 (b,and f) shows enthalpy for BeS, BeSe and BeTe respectively. When the enthalpy goes down as pressure increases for BeS, BeSe and BeTe, it shows that the material shrinks and its structure rearranges. It is clear this type of behavior points to a shift into a denser form with more order, mainly in BeS because it has stronger bonds and smaller ions. This means pressure helps reduce the effort needed to turn gases into a solid and holds the atomic structure together more firmly. A decrease in enthalpy with increasing pressure at constant temperature indicates that the system is becoming more stable in a denser configuration. This implies a contraction in molar volume and may correspond to a negative or reduced volumetric expansion coefficient, highlighting the compressible nature of the material under applied pressure.

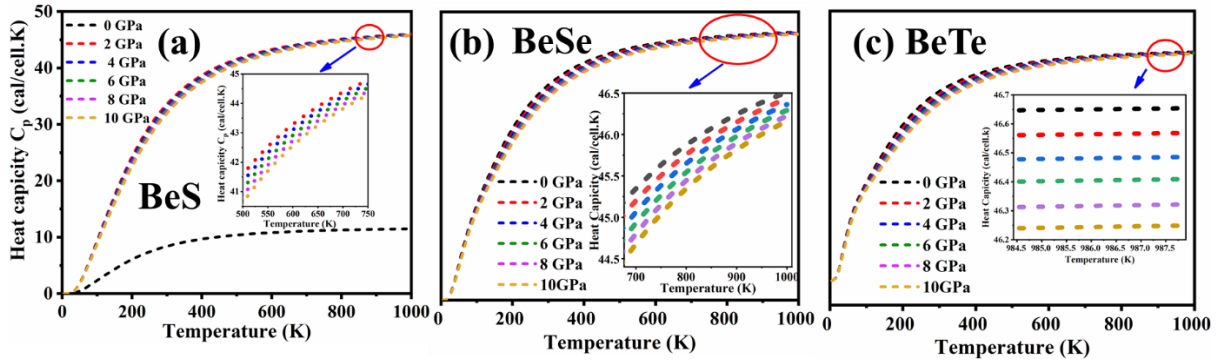

**Fig.17.** Illustration of heat capacity of (a) BeS (b) BeSe and (c) BeTe under the effect of different pressures.

In figure 17 (a-c) explain how the heat capacity is constant when pressure stays the same. These properties in BeS, BeSe and BeTe change as temperature varies at different pressures (0-10 GPa). The capacity to hold heat gets bigger as the temperature goes up for BeS, BeSe and BeTe compounds. This happens due to the fact that as temperature increases, more energy-absorbing methods (phonon modes) are working in the system. As temperatures rise (~1000 K), the heat capacity of metals closes in on the Du-long-Petite value (approximately 3R or 25 J/mol·K per atom), as expected for solids by classical theory [38, 39]. Below 300 K, $C_p$ exhibit an upward trend, though their rate of increase becomes more gradual at elevated temperatures.

**Table .2** calculated thermodynamic parameters Gibbs energy (G), entropy (S), enthalpy (H), Debye temperature ($\theta_D$), heat capicity ($C_p$) and molar mass (M).

| Compounds | P (Gpa) | G (eV) | S(eV /K) | H (eV /atom) | θ D(K) | Cp (cal /cell. K) | M (a.m.u) |
|---|---|---|---|---|---|---|---|
| BeTe | 0 | -2.6967 | 1.0907 | 4.3283 | 623.24 | 46.65 | 135.61 |
|  | 2 | -2.6566 | 1.0788 | 4.2721 | 657.01 | 46.55 |  |
|  | 4 | -2.6143 | 1.0675 | 4.2207 | 679.27 | 46.47 |  |
|  | 6 | -2.5795 | 1.0574 | 4.1762 | 701.52 | 46.39 |  |
|  | 8 | -2.5434 | 1.0473 | 4.1231 | 725.80 | 46.31 |  |
|  | 10 | -2.5141 | 1.0390 | 4.0872 | 746.03 | 46.23 |  |
| BeSe | 0 | -2.3230 | 3.7869 | 1.4190 | 685.61 | 45.11 | 86.97 |
|  | 2 | -2.2843 | 3.9035 | 1.4083 | 707.13 | 25.17 |  |
|  | 4 | -2.2592 | 3.8632 | 1.3977 | 726.99 | 45.02 |  |
|  | 6 | -2.2302 | 3.8263 | 1.3883 | 745.20 | 44.88 |  |
|  | 8 | -2.2074 | 3.7904 | 1.3801 | 761.75 | 44.76 |  |
|  | 10 | -2.1826 | 3.7580 | 1.3715 | 776.65 | 44.63 |  |
| BeS | 0 | -0.4550 | 824.36 | 0.3766 | 831.86 | 11.4535 | 40.08 |
|  | 2 | -1.8252 | 833.13 | 1.4923 | 832.78 | 44.8023 |  |
|  | 4 | -1.8000 | 853.36 | 1.4815 | 851.17 | 44.6306 |  |
|  | 6 | -1.7788 | 87.62 | 1.4728 | 868.64 | 44.4269 |  |
|  | 8 | -1.7558 | 886.33 | 1.4637 | 885.20 | 44.3210 |  |
|  | 10 | -1.7380 | 903.80 | 1.4555 | 899.55 | 44.1798 |  |

This behavior suggests that ion-ion interactions in compounds strongly influence heat capacity in the low-temperature range. As the temperature continues to rise, the curve approaches linearity, aligning with the predictions of the Du-long-Petite law. While both temperature and pressure affect heat capacity, temperature plays a more dominant role, particularly below 300 K, where its influence outweighs that of pressure. The order of $C_p$ from highest to lowest BeTe > BeSe > BeS. BeS. The BeTe and BeSe have a higher heat capacity at higher temperatures because there are more phonon movements. Pressure pushing on an atom decreases the energy needed to vibrate by compressing the lattice and reducing heat capacity. BeS shows the most heat-handling capability which is due to its strict and small structure.

## 4. Conclusions

This study thoroughly investigated the structure, electronic, optical and thermodynamic behavior of BeS, BeSe and BeTe, using first-principles density functional theory (DFT) under hydrostatic pressure ranging from 0 to 10 GPa. The analysis shows that the results are agreed with previous studies. At all pressures up to 10 GPa, BeS, BeSe and BeTe remain in the zinc blende (cubic) structure and do not transform to any other phase. Because all three compounds can resist compression, they have outstanding thermal and structural stability. The crystals lattice are compressible with BeS being the least compressible, while BeSe and BeTe exhibit higher compressibility. This kind of response means that elastic constants and bulk modulus increase which demonstrates the materials' strong mechanical qualities. Applying pressure significantly alters the electronic band structures of the materials, leading to a decreases in band gap for each compound. Among them BeS consistently-exhibit the widest band-gap. It is observed from the optical absorption spectra that the absorption starts in the visible region and peak in the UV range, which shows that the material has strong optical properties and can be used in UV optoelectronics applications. The presence of positive phonon frequencies throughout the Brillouin zone for BeS, BeSe and BeTe confirm the materials are dynamically stable under all simulated pressures. The fact that electron temperature increases as pressure increases it indicates that electrons have more energy and this should help improve the thermoelectric performance and conductivity of these materials when they are subjected to high pressures.

Therefore, the results for BeS, BeSe and BeTe reveals that they hold significant potential for future optoelectronic gadgets, advancements in high-pressure tools and low-energy applications such as white LED's, UV sensors and thermoelectric devices. Understanding their behavior under varying pressures provides a strong foundation for their potential use in thermoelectric devices, optoelectronics, and thermal barrier coatings. The findings obtained by computation provide new chances for further observing and testing beryllium chalcogenides in various kinds of engineering.

**Declaration of competing interest**

The authors state that financial and personal conflicts of interest did not influence the outcomes and summary reached in this study.

**Data availability**

Data will be made available on request

**Author contributions statement: Muhammad Shahzad**: Data Curation, Writing-Original Draft. **Ming Li**: Conceptualization, Supervision, investigation and funding acquisition. **Sikander Azam**: Writing-Reviewing and Editing. **Syed Awais Ahmad:** Analysis and methodology.

**Acknowledgments:**

This work was funded by the Major Science and Technology Special Project of Yunnan Province (Grant No. 202202AE090011) and the Basic Research Program of Yunnan Province (Grant No. 202301AT070082).